\documentclass[12pt,preprint]{aastex}
\usepackage{lscape}
\usepackage{epsfig}
\usepackage{url}
\newcommand{\psr}{PSR~B1259$-$63}
\newcommand{\sta}{LS~2883}
\newcommand{\binary}{PSR~B1259$-$63/LS~2883}
\newcommand{\gr}{$\gamma$-ray}
\newcommand{\grs}{$\gamma$-rays}
\newcommand{\flux}{\,erg\,cm$^{-2}$\,s$^{-1}$}

\newcommand{\cm}{\,cm$^{-2}$}
\newcommand{\nh}{$N_\mathrm{H}$}

\begin{document}

\title{High-energy observations of \binary~through the 2014 periastron passage: connecting X-rays to the GeV flare}

\author{P.~H.~T. Tam$^{1,2}$, K.~L. Li$^{1}$, J. Takata$^{2}$, A.~T. Okazaki$^{3}$, C.~Y. Hui$^{4}$, A.~K.~H. Kong$^{1}$}
\affil {$^1$ Institute of Astronomy and Department of Physics, National Tsing Hua University, Hsinchu, Taiwan\\
$^2$ Department of Physics, University of Hong Kong, Pokfulam Road, Hong Kong\\
$^3$  Faculty of Engineering, Hokkai-Gakuen University, Toyohira-ku, Sapporo 062-8605, Japan\\
$^4$ Department of Astronomy and Space Science, Chungnam National University, Daejeon, Republic of Korea
}
\email{phtam@phys.nthu.edu.tw}

\begin{abstract}
The binary system \binary~is well sampled in radio, X-rays, and TeV $\gamma$-rays, and shows orbital-phase-dependent variability in these frequencies. The first detection of GeV $\gamma$-rays from the system was made around the 2010 periastron passage. In this Letter, we present an analysis of X-ray and \gr~data obtained by the Swift/XRT, NuSTAR/FPM, and Fermi/LAT, through the recent periastron passage which occurred on 2014 May 4. While \binary~was not detected by the LAT before and during this passage, we show that the GeV flares occurred at a similar orbital phase as in early 2011, thus establishing the repetitive nature of the post-periastron GeV flares. Multiple flares each lasting for a few days have been observed and short-term variability is seen as well. We also found X-ray flux variation contemporaneous with the GeV flare for the first time. A strong evidence of the keV-to-GeV connection came from the broadband high-energy spectra, which we interpret as synchrotron radiation from the shocked pulsar wind.

\end{abstract}


\section{Introduction}

\binary~is a unique binary system in our Galaxy located at $d=2.3\pm0.4$~kpc away~\citep{negueruela_astro_par}. It comprises the young radio pulsar, \psr, with a period 47.8~ms and the fast-rotating massive star \sta~\citep{1259_radio_94}. Moving in an eccentric ($e\sim0.87$) orbit, the pulsar passes the periastron every 1236.7 days~\citep[or 3.4 years;][]{orbital_shannon_14}. \sta~hosts a stellar disk that is inclined with respect to the orbital plane, such that the pulsar crosses the disk plane just before and just after each periastron passage~\citep{Wex98}. The pulsar generates a powerful pulsar wind (PW) that interacts with the stellar wind/disk, emitting broadband non-thermal, unpulsed radiation, in radio~\citep[e.g.,][]{1259_radio_05}, X-rays~\citep[e.g.,][]{Chernyakova_09}, and TeV $\gamma$-rays~\citep{hess_1259_05,hess_1259_09}. Because of the highly-eccentric orbit, the broadband radiation varies over the 3.4-year orbit, and peaks around the periastron passages~\citep[e.g., see][]{Chernyakova_06,Chernyakova_09}.

Radio imaging observations have revealed variable spatially-extended emission whose size is larger than the binary orbit~\citep{extended_radio_moldon_11}. Chandra observations near apastron also show extended X-ray emission~\citep{chandra_pavlov_11}.

Over the last periastron passage which occurred December 2010, the most surprising behavior came from the GeV band. It was the first Fermi Large Area Telescope (LAT) observations of the system through a periastron passage. Before and during the passage, the LAT detected a weak emission above 100~MeV, before, quite surprisingly, a GeV flare occurred $\sim$30 days after the passage~\citep{Kong_1259_atel}, with a flux about 10 times the pre-periastron value. The flares continued until about three months after the periastron passage~\citep{Tam_1259_2011,Abdo_1259_2011}.

H.E.S.S. observations of the binary were performed just before and during part of the GeV flaring period in January 2011. No significant brightening above 1~TeV was seen, and the flare coefficient, $\kappa_\mathrm{TeV}$, i.e., the $>$1~TeV flux ratio during the flare and before the flare, is constrained to be $\kappa_\mathrm{TeV}<3.5$ at the 99.7\% confidence level~\citep{HESS_1259_2011}. No X-ray flare was seen during the GeV flare as well~\citep[e.g.,][]{Chernyakova_14}.

The radiation mechanisms of the broadband radiation from \binary, as well as the GeV flare seen in early 2011, are unclear. Electrons accelerated in the shock between the PW and stellar wind can produce synchrotron radiation and/or upscatter stellar photons from \sta~to produce inverse-Compton (IC) radiation~\citep{Tavani97,Kirk99,Dubus06,Bogovalov08,Khangulyan11,Kong_model_11,Mochol13,Takata09,Takata12}. The unshocked PW particles may also generate \grs~\citep{Khangulyan12}. The interaction between the stellar disk and the pulsar~\citep{Chernyakova_14}, as well as Doppler boosting~\citep{Dubus10,Kong_model_12} may also play a role. The extended radio emission around periastron may be generated by particles that have traveled outside the binary system. High-energy observations during the 2014 periastron passage may help us to distinguish competing scenarios.

In this Letter, Fermi/LAT, NuSTAR/FPM, and Swift/XRT analysis results of the binary \binary~over the periastron passage which occurred on 2014 May 4 (or, $t_\mathrm{p}=MJD 56781.4195$) are presented.

\section{Gamma-ray Observations and Results}
\subsection{Observations and data analysis}
The $\gamma$-ray data\footnote{provided by the FSSC at \url{http://fermi.gsfc.nasa.gov/ssc/}} used in this work were obtained using the Fermi/LAT~\citep{lat_technical} between 2008 August 4 and 2014 August 2. In response to Target-of-opportunity (ToO) requests, observations targeted at \binary~were commenced from 2014 May 31 to June 26, such that the exposure toward \psr~was increased during the above period. We used the Fermi Science Tools v9r32p5 package to reduce and analyze the data. Reprocessed Pass 7 data classified as ``source'' events were used. To reduce the contamination from Earth albedo $\gamma$-rays, we excluded events with zenith angles greater than 100$^\circ$. The instrument response functions ``P7REP\_SOURCE\_V15'' were used.

We first carried out a binned maximum-likelihood analysis (\emph{gtlike}) of a rectangular region of 21$^\circ\times$21$^\circ$ centered on the position of \binary, using 6-year data. We subtracted the background contribution by including the Galactic diffuse model (gll\_iem\_v05.fit) and the isotropic background (iso\_iem\_v05.txt), as well as the second Fermi/LAT catalog~\citep[2FGL;][]{lat_2nd_cat} sources within 25$^\circ$ away from \psr. The \gr~source associated with the supernova remnant Kes~17 was also included~\citep{Jason_kes17}. The recommended spectral model for each source as in the 2FGL~catalog was used, while we modeled \psr~and Kes~17 with a power law (PL)
\begin{equation}
\frac{dN}{dE} = N_0 \left(\frac{E}{E_0}\right)^{-\Gamma}.
\end{equation}
The spectral values of \psr~and other sources within 5$^\circ$ from \psr, as well as normalization parameter values for the Galactic and isotropic diffuse components, and sources in the annulus of inner and outer radii of 5$^\circ$ and 10$^\circ$ from \psr, were allowed to vary. Other parameters were fixed.

\binary~was detected with TS=32.8 in this analysis. The source model thus obtained was used as a template for analyzes hereafter. The normalization parameter values for \psr, the Galactic diffuse component, and sources within 4$^\circ$ from \psr~and the two sources marked as variable (i.e., 2FGL~J1329.2$-$5608 and 2FGL~J1330.1$-$7002) in the 2FGL catalog, were allowed to vary, in subsequent analyzes.

\subsection{Light curve}
We derived the 0.1--300~GeV light curve composed of 5-day bins (Fig. 1), assuming PL for \binary. The photon indices for bins with TS$>$5 are presented as well. The light curve around the 2010 passage are shown for comparison.

As shown in Fig. 1, no significant emission was found in any time intervals before and during the passage on 2014 May 4, except for the single bin with $TS=9.6$ centered at $t_\mathrm{p}-$17.5~days. The source only became significantly detected from 2014 June 6 (i.e., $t_\mathrm{p}+$33d) on~\citep{Atel6216,Atel6225}, similar to the orbital phase when the last GeV flare was observed in 2011. The major flaring period continues up to around $t_\mathrm{p}+60$~days.

To probe shorter time-scale variability during the flares, we also derived 0.1--300~GeV light curves composed of 1-day/12-hr bins (Fig.~\ref{zoomin_plot}), fixing the PL index of \psr~at 3.0 (the average photon index during the 2014 flare period).

The \gr~daily light curve is highly variable from $t_\mathrm{p}+$30d to $t_\mathrm{p}+$60d, clearly consisting of multiple flares. Three peaks are identified at 38d, 49d, and 57d after $t_\mathrm{p}$, lasting for 2 to 6 days. Using 12-h bins, it can be seen that \grs~from \binary~undergo rapid variations down to time scale of 12 hours. Variations at time scales down to three hours are also seen.

\subsection{Spectral analysis during the 2014 flaring period}
The GeV flare formally started on 2014 June 6, but in light of the X-ray hardness revealed by the NuSTAR data (see Table~\ref{joint_xray_spec}), we regard the flaring period to be between June 2 and July 2 in this section, and the corresponding \gr~spectrum is shown in Fig.~\ref{sed}. Flux values of the 9 energy bin were reconstructed using \emph{gtlike} for each band independently, using a representative photon index of $\Gamma_\gamma=$3 for each bin. The PL with an exponential cutoff (PLE) model \begin{equation}
\frac{dN}{dE} = N_0 \left(\frac{E}{E_0}\right)^{-\Gamma}\mathrm{exp}\left(-\frac{E}{E_\mathrm{c}}\right).
\end{equation}
describes the 0.1--300 GeV spectrum better than PL by $\Delta$TS$=$20.4, i.e.,$\sim$4.5$\sigma$ in significance (see Table~\ref{flares}).

We also defined three flaring episodes: (f1) June 6 to June 16; (f2) June 17 to June 25; and (f3) June 26 to July 2. We found that a PLE describes the spectrum better than a PL during f1, with $\Delta$TS$=$60.5, i.e., $\sim$7.8$\sigma$ in significance. The best-fit parameters are shown in Table~\ref{flares}.

The 50--300~GeV bin was detected with TS=8.3 during the flaring period, and is dominated by a 50~GeV photon which arrived on 2014 June 10 and is located $2\farcm4$ from \psr. We speculate that this photon is related to the spectral component in the TeV band at similar orbital phase~\citep{HESS_1259_2011}.

\subsection{Comparisons with the 2010 periastron passage}
\label{compare}
Unlike the previous periastron passage, \binary~was not detected by the LAT before and during the 2014 passage. This may not be too surprising given the marginal detection last time. This shows that increasing the exposure before and during the next periastron passages is very important.

The GeV flares occurred at similar orbital phase as in early 2011, clearly demonstrating the repetitive nature of the GeV flares after the periastron passages in 2010 and 2014. Multi-peaked structure in the light curve was first seen in 2011~\citep[see Fig.~2 of][]{Tam_1259_2011}, and is seen again this time. Although during the first flare the flux rises at a rate slower than in January 2011, the first peak occur at nearly the same orbital phase ($t_\mathrm{p}+$37d and $t_\mathrm{p}+$38d in 2011 and 2014, respectively).

\section{X-ray Analysis and Results}

A \textit{Swift}/XRT monitoring campaign consisting of 29 observations was performed, resulting in a total exposure of 70~ks spanning from April 20 ($t_\mathrm{p}-14$d) to July 8 ($t_\mathrm{p}+65$d). As the observed flux is too high during some of the observations, Windowed Timing (WT) mode, instead of the conventional Photon Counting (PC) mode, was partially used to avoid the pile-up problem. For WT data, there is an artificial turn-up in the low energy band (i.e., $<$ 1.0~keV), which is a known calibration issue of this mode\footnote{\url{http://www.swift.ac.uk/analysis/xrt/digest\_cal.php}}. Therefore, we excluded data below 1.0~keV for WT data throughout the X-ray analysis, while the full band of \textit{Swift}/XRT (i.e. 0.3--10~keV) was used for PC-mode observations.

Five NuSTAR observations were taken from April to June 2014 with a total exposure of 150~ks, which are sensitive to 3--79~keV photons.

We downloaded the \textit{Swift}/XRT and NuSTAR/FPM data from the HEASARC data archive, and extracted scientific products for each observation using the tasks \texttt{HEAsoft} 6.14 \texttt{xrtgrblc} and \texttt{xrtgrblcspec}. Count rate-dependent regions were generated by \texttt{xrtgrblc} for the XRT data and by \texttt{nuproducts} with 30$\arcsec$ radius source region and 130$\arcsec$ source-free background region for the FPM data. We fit the extracted X-ray spectra with \texttt{XSPEC} 12.8.1.

For \textit{Swift} data, because of the limited quality of individual spectrum and the lack of soft X-rays for WT data, absorption level of the source is hard to estimate.
We therefore splitted the observations into two groups (i.e., pre- and post-periastron) and set each group a common column density (\nh) for better \nh\ constraint.
Although the interaction of \binary\ winds may introduce small influence on \nh, the Galactic column density towards \binary\ is as high as $10^{22}$ \cm\ \citep{Kalberla05} that is comparable with the best-fit \nh\ $\approx 5-7\times10^{21}$ \cm\ of this source (see below), indicating that the absorption is dominated by the Galactic foreground gas and the intrinsic \nh\ variation of the system is negligible. All the spectra are well fit with an absorbed power-law with $\chi_\nu^2$ values of 0.80 (dof=34; pre-periastron) and 0.98 (dof=674; post-periastron). The \nh\ value before/after the passage is $6.9^{-0.08}_{+0.09}\times10^{21}$\cm\ and $5.4^{-0.04}_{+0.05}\times10^{21}$\cm, respectively. The photon index keeps changing throughout the periastron passage, reaching $2.06^{+0.18}_{-0.17}$ on April 20, gradually dropping to $1.22^{+0.21}_{-0.21}$ on June 24, and staying around at 1.5 afterward. Indeed, fitting the photon indices with a linear function gives $\Gamma_\mathrm{XRT}=(1.89\pm0.04)-(0.006\pm0.001)[t(d)-t_\mathrm{p}]$ (see Fig. 1). The decreasing trend is confirmed with the F-test probability of $1.6\times10^{-6}$. This hardening of X-ray emission is further established by NuSTAR observations. Previous X-ray observations showed similar X-ray spectral hardening through periastron passages~\citep[e.g.,][]{Chernyakova_09,Chernyakova_14}, and this behavior now extends to 79~keV thanks to NuSTAR observations.

The X-ray flux variation is compatible with the previously observed double-hump structure~\citep{Chernyakova_09}.
Although no huge flare is detected, there is seemingly short-term variability along the PL-like declining trend from 2014~June~2 to July~8 (panel(e) of Fig. 1). X-ray variability of short time scales, i.e., down to $\sim$1 ks, was previously reported in two epochs of pre-periastron observations ($t_\mathrm{p}-$19d and $t_\mathrm{p}-$11d), attributed to the clumpy structure of the stellar wind~\citep{Chernyakova_09}. With data taken from the GeV-flaring period June~2 to July~8, the best-fit decay index of the trend is $-0.47\pm0.01$ with a $\chi_\nu^2=3.3$ (dof=16). As the model matches well with the declining trend, we suggest that the large $\chi^2$ value is indeed caused by short-term variability. If so, the significance of short-term variability is more than $5\sigma$. We speculate that the variability is mainly caused by several small-scale flaring events. We therefore construct a baseline light curve, $F\propto (t-t_0)^{-0.47}$ where $t_0$ is the observed highest peak of the light curve (see panel (e) of Fig. 1; the X-ray peak at $t_0$ has the highest flux ever detected from \binary, first noted in Tam and Kong, 2014), based on the three low points of the phase (that occurred on June 2, June 14, and July 8) to investigate the amplitude of the flares. By defining a flare coefficient, $\kappa_\mathrm{X}$, as the ratio between the observed flux (absorption corrected) and the baseline model flux (i.e., $\kappa_\mathrm{X}=F_\mathrm{observed}/F_\mathrm{baseline}$), we show that the nominal flare occupies a considerable fraction of the total flux, equivalent to 10\%--60\% of the baseline flux. The flare coefficient during the GeV-flaring period is about $\kappa_\mathrm{X}=$1.1 to $\kappa_\mathrm{X}=$1.6.

We fit the NuSTAR spectra simultaneously with the corresponding \textit{Swift}/XRT data and all the combined spectra can be described by an absorbed power-law (overall $\chi^2_\nu=0.98$ with dof=3671; the best-fit photon indices and fluxes are shown in Table~\ref{joint_xray_spec}). In Fig.~\ref{sed}, we plotted three spectra using contemporaneous XRT+FPM spectra (i.e., phase I: April 20 to May 14; phase II: May 15 to June 1; and phase III: June 2 to July 2), representing the periastron passage, X-ray peak, and flaring phases, respectively.

\section{Discussion}
 \citet{Kong_model_12} proposed that the GeV flare is due to Doppler boosting of the shocked PW. In this model, the line of sight cuts through the shocked PW cone during the phase of the GeV flares, boosting the emissions through relativistic effects. However, because the Doppler boosting not only affects the GeV band but also the X-ray band, the peak of the GeV light curve is in phase with the post-periastron X-ray peak, which is incompatible with the observed GeV/X-ray light curves (c.f., Fig. 1). The inverse-Compton scattering process of the PW has been discussed before~\citep{Soelen12, Khangulyan12,Dubus13,Mochol13},
while different seed photons are assumed in different models.
\citet{Soelen12} 
calculated the inverse-Compton scattering process between the PW 
and infrared photons from the Be disk. In the \gr~light curve, 
the disk contribution is a maximum at the periastron passage, because the disk 
is the brightest near the stellar surface.  \citet{Dubus13} discussed the inverse-Compton scattering of the PW off the X-rays from the shocked PW. However, this model does not explain the observed temporal delay between the X-ray peak and GeV flare. \citet{Khangulyan12} proposed that
the GeV flare occurs as a consequence of the rapid changing of shock when the pulsar exits the stellar disk. The opening  of the shock cone enables the cold-relativistic PW to travel further along the line of sight, and to cool through the inverse-Compton scattering off the optical photons from the disk. This model can explain the delay between the GeV flare and the X-ray peak. In this model,  a large energy density of the seed photons reaching about 10 times greater than the stellar radiation density is required~\citep[c.f.,][]{Dubus13}.
However, observations through previous periastron passages do not seem to suggest such a large heating effect (Negueruela et al. 2011)
 and dedicated optical and IR observations during the {\it Fermi}
 flare  are required to investigate the heating of the disk matter 
by the PW. A hydrodynamic simulation (Takata et al. 2012) suggests
that the disk matter is deformed by the pulsar and creates a cavity structure 
around the pulsar. The cavity structure may allow the head-on collision 
between the PW and the IR photons from the heated matter, increasing 
the efficiency of \gr~production. 

As shown in Table~\ref{joint_xray_spec} and Fig. 1, new X-ray observations indicate the hardening of spectra up to 79~keV over time, and the linear extrapolation of the X-ray spectrum to the high-energy band roughly connects to the Fermi data during phase III (c.f. Fig.~\ref{sed}), suggesting that X-ray and GeV emissions are related to each other, and GeV flare emission likely originates from the synchrotron radiation processes. \citet{Chernyakova_14} proposed a
synchrotron radiation scenario, in which the fly-by of the disk material disturbs the magnetic field of the un-shocked PW and  produces the pitch angle of the PW particles, which can be cooled down through the synchrotron radiation. 

In Fig.~\ref{sed}, we compare our calculated spectra of the synchrotron radiation from the shocked PW with the observed spectra before and during the GeV flares. We used the shock structure obtained by a three-dimensional smoothed particle hydrodynamics simulation for the Be-disk base density of $10^{-9}$~g~cm$^{-3}$ (Okazaki et al., in prep.). This simulation has been carried out using the same model as in~\citet{Takata12}, but with updated stellar and wind parameters~\citep{negueruela_astro_par,extended_radio_moldon_11}.
 We assumed that the PW is accelerated  at the shock and the maximum Lorentz factor is determined by balancing 
the acceleration time scale ($\gamma m_ec/eB$) with the synchrotron loss (or dynamical) time scale. The typical Lorentz factor in this work is of the order
$\gamma_{max}\sim 10^8$  and makes a spectral break of the synchrotron 
radiation at 10--100~MeV.  
We  chose the power index, $p$, of the shocked particles to explain the 
X-ray spectral index, and we assumed $p=2.7$ for phase I, $p=2.4$ for phase II, and $p=2.0$ for phase III. In calculating the 
synchrotron emission with the result of hydrodynamic simulation, we assumed that the total PW pressure at 
each simulation grid is associated with 20\% of the magnetic pressure and 80\% of the particle pressure.  The typical magnetic field at the shock region in this work is $B\sim 0.05-1$~Gauss.   We found that if 
the magnetic field at the shock is $B\ll 0.1$~Gauss,  the calculated flux is well below the observed flux. 
 We refer to \citet{Takata12} for details of the calculations. This model interprets the pre- and post-periastron X-ray peaks as due to significant increase of the conversion efficiency from pulsar spin-down power to the shock-accelerated particle energy at orbital phases when the pulsar crosses the disk.

\acknowledgments
We acknowledge the use of data and software facilities
from the FSSC, managed by the HEASARC at the Goddard Space Flight Center. PHT and KLL are supported by the Ministry of Science and Technology (MOST) of the Republic of China (Taiwan) through grant 101-2112-M-007-022-MY3. AKHK is supported by MOST through grants 100-2628-M-007-002-MY3, 100-2923-M-007-001-MY3, and 103-2628-M-007-003-MY3. JT is supported by a 2014 GRF grant of Hong Kong Government under HKU 17300814P. CYH is supported by the Chungnam National University research fund in 2014.

\begin{landscape}
\begin{table}
\begin{center}
\caption{0.1--300~GeV spectral properties during the 2014 flaring period \label{flares}}
\begin{footnotesize}
\begin{tabular}{l@{ }clccc@{ }c@{ }r@{ }r}
    \tableline\tableline
    Period & Date & Model & Photon Flux & Energy Flux
    & Photon Index & Cutoff Energy & TS & TS$_\mathrm{cutoff}$ \\
           &     &   & (cm$^{-2}$~s$^{-1}$) &(erg~cm$^{-2}$~s$^{-1}$) &           & (MeV)     & & $\mathrm{TS}_\mathrm{PLE}-\mathrm{TS}_\mathrm{PL}$ \\
    \tableline
    f1  & June 6 -- June 16 & PL & (1.16$\pm$0.09)$\times$10$^{-6}$ & (3.59$\pm$0.29)$\times$10$^{-10}$
    & 3.07$\pm$0.10 & & 318.0 & \\
       &                    & PLE & (1.17$\pm$0.07)$\times$10$^{-6}$ & (3.50$\pm$0.27)$\times$10$^{-10}$ & 2.12$\pm$0.41 & 331$\pm$164 & 378.5 & 60.5 \\
    f2  & June 17 -- June 25 & PL & (7.28$\pm$1.32)$\times$10$^{-7}$ & (2.61$\pm$0.42)$\times$10$^{-10}$ & 2.63$\pm$0.13 & & 84.0 & \\
       &                     & PLE & (6.91$\pm$0.14)$\times$10$^{-7}$ & (2.76$\pm$0.45)$\times$10$^{-10}$ & 2.26$\pm$0.31 & 1899$\pm$1810 & 83.4 & $^a$ \\
    f3  & June 26 -- July 2 & PL & (1.20$\pm$0.22)$\times$10$^{-6}$ & (4.04$\pm$0.73)$\times$10$^{-10}$ & 2.90$\pm$0.17 & & 88.7 & \\
           &                & PLE & (1.14$\pm$0.24)$\times$10$^{-6}$  & (3.95$\pm$0.70)$\times$10$^{-10}$
           & 2.37$\pm$0.44 & 1066$\pm$914 & 87.4 & $^a$ \\
    flare  & June 2 -- July 2 & PL & (9.16$\pm$0.76)$\times$10$^{-7}$ & (3.10$\pm$0.24)$\times$10$^{-10}$ & 2.90$\pm$0.07 & & 459.1 & \\
        &                 & PLE & (9.06$\pm$0.77)$\times$10$^{-7}$  & (2.94$\pm$0.22)$\times$10$^{-10}$
        & 2.30$\pm$0.25 & 653$\pm$298 & 479.5 & 20.4 \\
    \tableline
\end{tabular}
\end{footnotesize}
\tablenotetext{a}{No improvement of the PLE model over the PL model.}
\end{center}
\end{table}
\end{landscape}

\begin{table}
\begin{center}
\caption{Joint XRT/FPM 0.3--79~keV spectral properties \label{joint_xray_spec}}
\begin{tabular}{l@{ }ccc}
    \tableline\tableline
     Date& \nh& $\Gamma$& Energy Flux\\
           & (cm$^{-2}$\,s$^{-1}$) &     & (\flux) \\
    \tableline
     April 20& 0.90$^{+0.13}_{-0.10}$ & 1.88$\pm$0.02 & (8.45$\pm$0.09)$\times10^{-11}$ \\
     May 4   & 0.69$^{+0.07}_{-0.06}$ & 1.94$\pm$0.02 & (4.51$\pm$0.06)$\times10^{-11}$ \\
     May 28  & 0.52$^{+0.09}_{-0.08}$ & 1.66$\pm$0.02 & (1.02$\pm$0.01)$\times10^{-10}$ \\
     June 2  & 0.67$^{+0.17}_{-0.15}$ & 1.57$\pm$0.02 & (9.14$\pm$0.12)$\times10^{-11}$ \\
     June 14 & 0.77$^{+0.25}_{-0.21}$ & 1.57$\pm$0.02 & (8.42$\pm$0.11)$\times10^{-11}$ \\
            \tableline
\end{tabular}
\end{center}
\end{table}

   \begin{figure}
\centerline{
\epsfig{figure=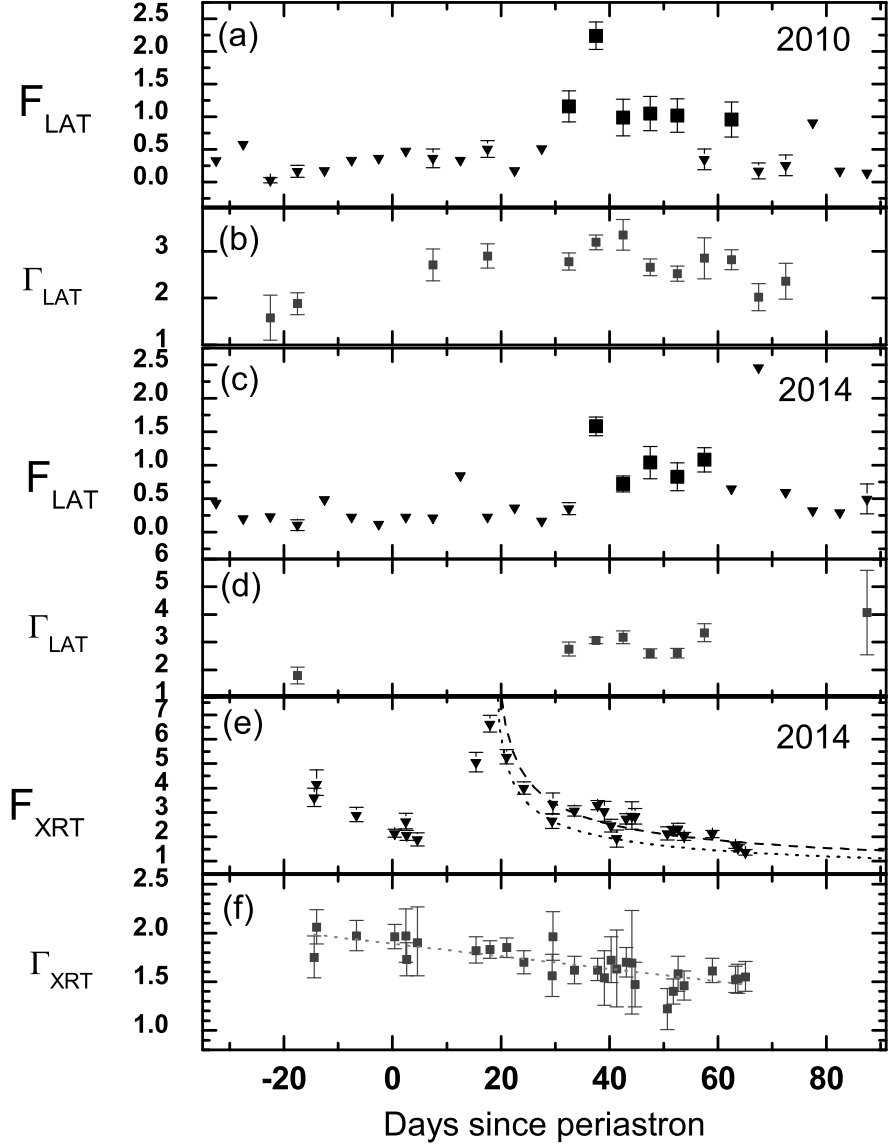,width=13cm}}
\begin{footnotesize}
      \caption{High-energy light curves of \binary\ over its recent periastron passages. (a) and (c): \emph{Fermi}/LAT 0.1--300~GeV light curve over the 2010 and 2014 periastron passage, respectively. Flux is given in $10^{-6}$\, ph\,cm$^{-2}$\,s$^{-1}$. Five-day time bins are used. Bins with TS$>$25 (significant) are indicated by squares. Bins with 5$<$TS$<$25 (marginally significant) are indicated by triangles. For bins with TS$<$5, 95\% confidence-level upper limits were calculated assuming $\Gamma=3$; (b) and (d): \gr~photon index evolution over the 2010 and 2014 periastron passage, respectively; (e) \emph{Swift}/XRT 0.3--10~keV light curve in 2014. Flux is given in $10^{-11}$~\flux. The best-fit baseline model (dotted) and the declining tend (dashed) are fitted using data after June 2. (f) X-ray photon index evolution in 2014. The dashed line shows the linear fit $\Gamma_\mathrm{XRT}=(1.89\pm0.04)-(0.006\pm0.001)[t(d)-t_\mathrm{p}]$.}
      \end{footnotesize}\label{5d_plot}
   \end{figure}

   \begin{figure}
\centerline{
\epsfig{figure=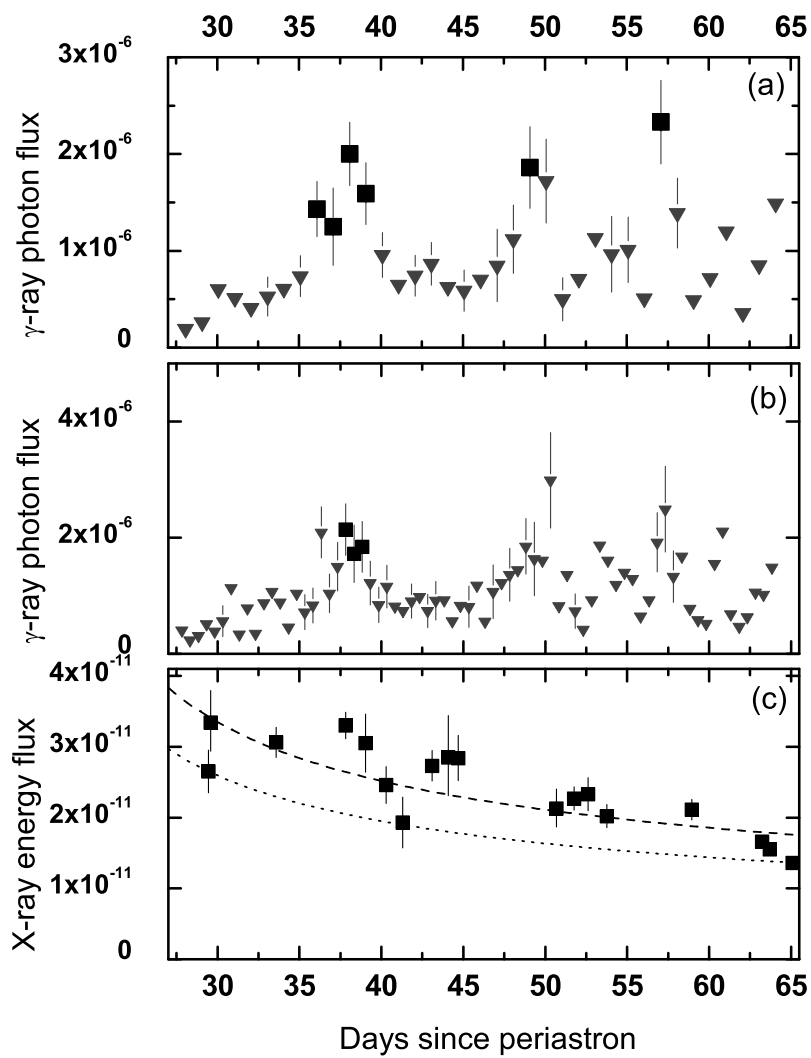,width=12cm}}
      \caption{Light curves of \binary\ during the 2014 GeV-flaring period. (a) and (b): Fermi/LAT 0.1--300~GeV light curve with daily and half-day bins, respectively. Symbols have the same meanings as in Fig.~1. (c) Swift/XRT 0.3--10~keV light curve. The dotted and dashed lines are the same ones shown in Fig.~1(e). }
      \label{zoomin_plot}
   \end{figure}

   \begin{figure}
\centering
\includegraphics[width=90mm]{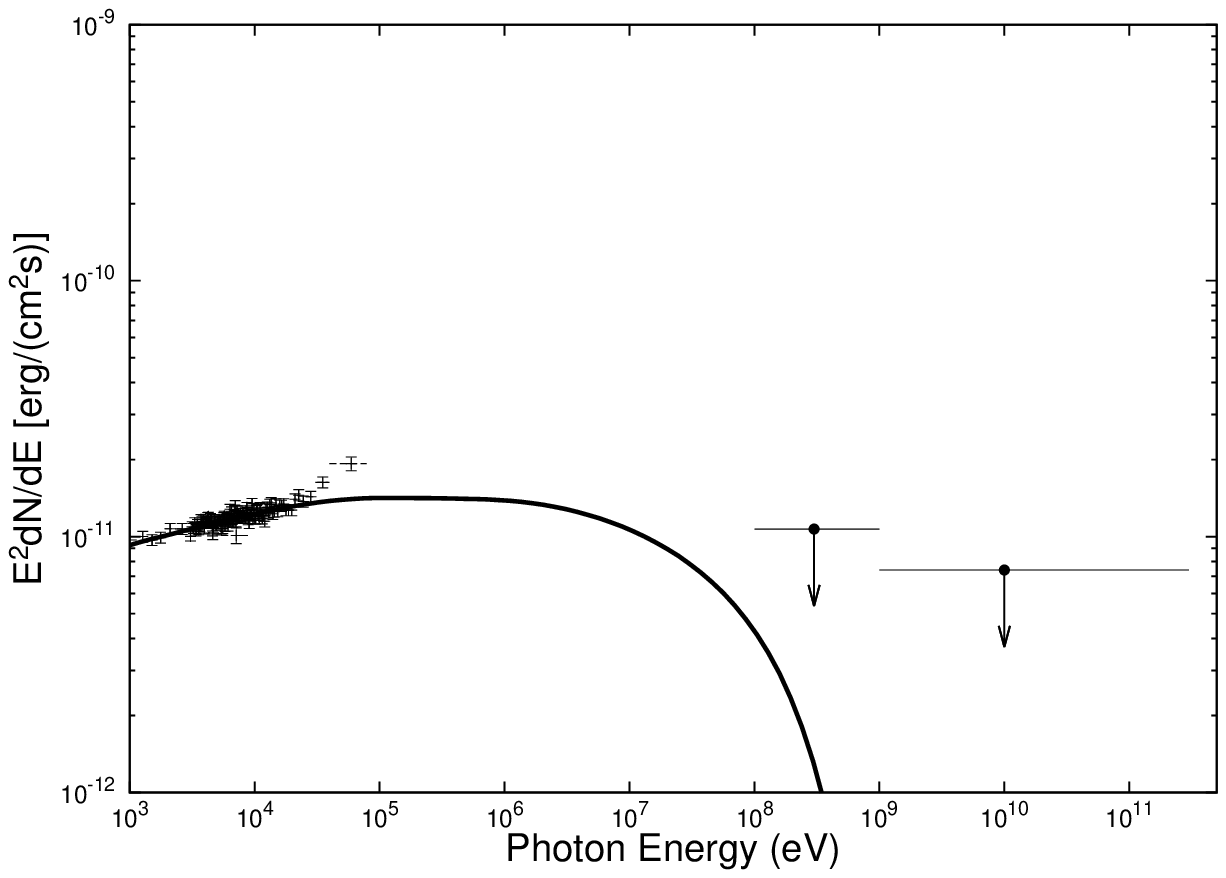}
\includegraphics[width=90mm]{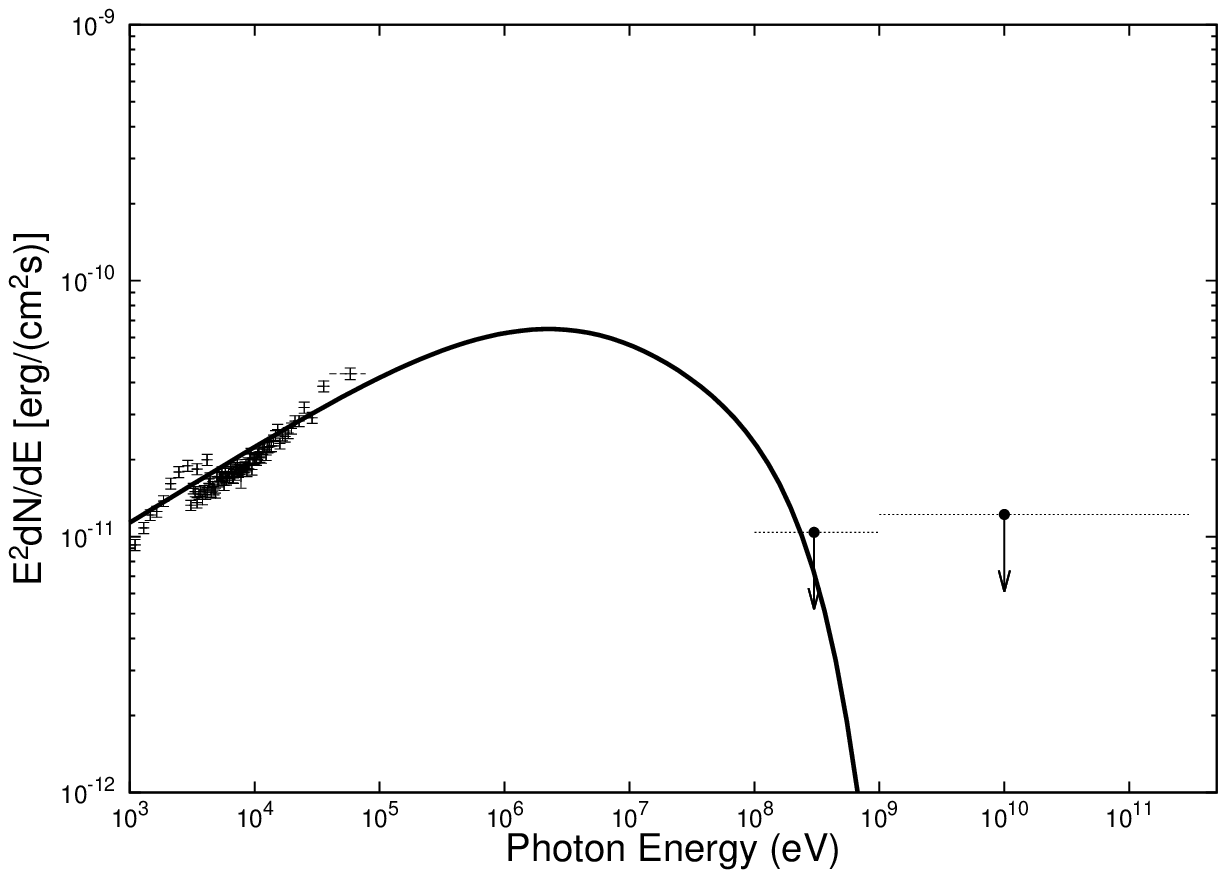}
\includegraphics[width=90mm]{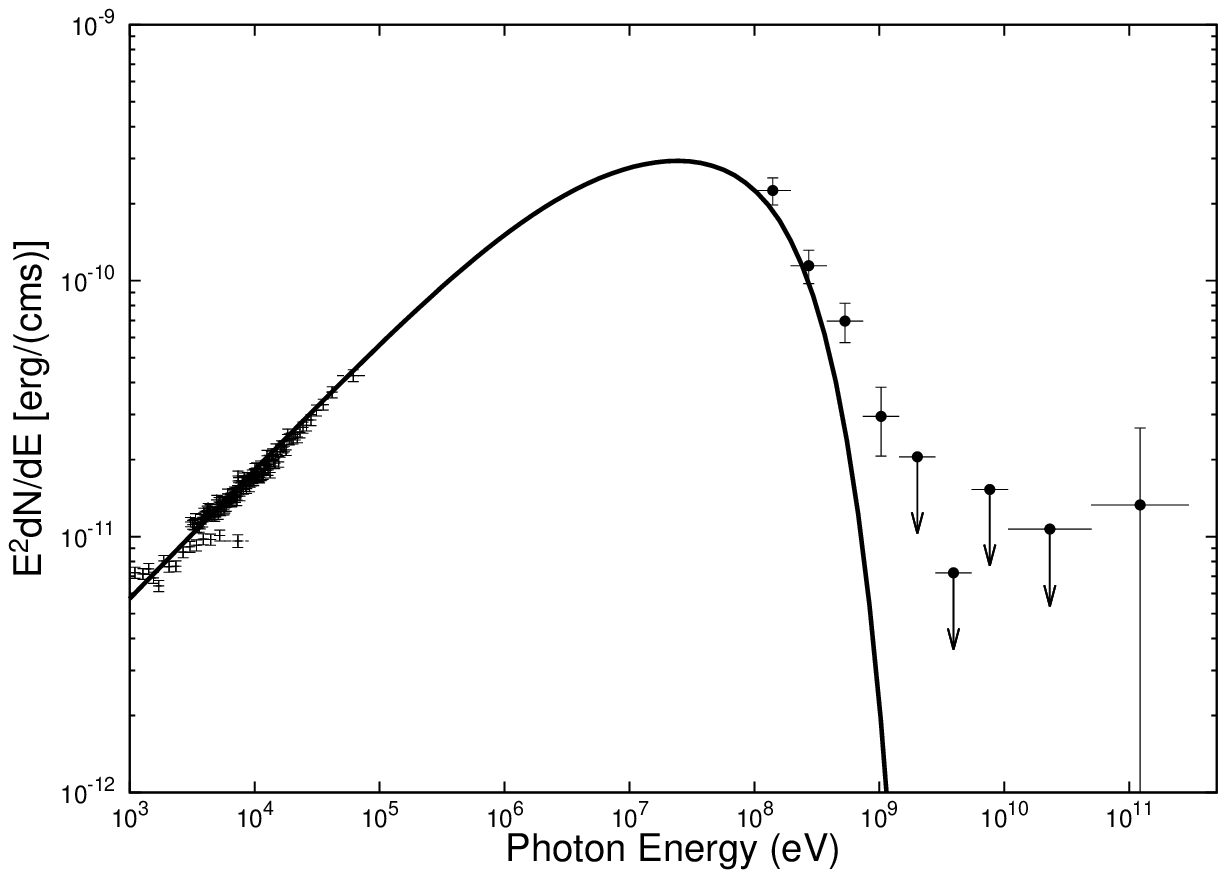}
      \caption{The spectral energy distribution from X-rays to \grs\ during phases I (periastron passage: April 20 to May 14; \emph{upper panel}), II (X-ray peak: May 15 to June 1; \emph{middle panel}), and III (flaring: June 2 to July 2; \emph{lower panel}). Some of the X-ray error bars are smaller than the symbols. Gamma-ray data without error bars represent 95\% confidence-level upper limits assuming $\Gamma=2$ for phases I and II, and $\Gamma=3$ for phase III. The solid lines show our calculated spectra of the synchrotron radiation from the shocked PW.}
      \label{sed}
   \end{figure}

\end{document}